# Tree diameter, height and stocking in even-aged forests

Jerome K VANCLAY
School of Environmental Science and Management
Southern Cross University
PO Box 157, Lismore NSW 2480, Australia
Tel +61 2 6620 3147
Fax +61 2 6621 2669
JVanclay@scu.edu.au

**Abstract**

Empirical observations suggest that in pure even-aged forests, the mean diameter of forest trees (*D*, diameter at breast height, 1.3 m above ground) tends to remain a constant proportion of stand height (*H*, average height of the largest trees in a stand) divided by the logarithm of stand density (*N*, number of trees per hectare): $D = \beta (H-1.3)/\text{Ln}(N)$. Thinning causes a relatively small and temporary change in the slope β, the magnitude and duration of which depends on the nature of the thinning. This relationship may provide a robust predictor of growth in situations where scarce data and resources preclude more sophisticated modelling approaches.

**Introduction**

Researchers have long sought a reliable relationship between tree size and stand density for use in forest inventory and management. Eichhorn (1904) observed that wood volume correlates well with stand height, and that this correlation is independent of stand age and site quality. However, the relationship between volume and stand height is a curve that may depend on stand management, restricting the utility of this relationship (Skovsgaard and Vanclay, 2008).

Other relationships between mean tree size and stand density have been suggested, and many of these can be generalised as $\ln N + \beta \ln S$ = constant, where *N* is stand density (stems/ha) and *S* is some measure of stem size (e.g., height, diameter or volume; Vanclay, 1994). For instance, Hart (1928; also Wilson, 1951) advocated relative spacing (*N*) which relates average spacing to top height (*H*): $\ln N + 2\ln H$ = constant; Reineke (1933) advocated a stand density index relating limiting stocking (*N*) to mean stem diameter (*D*): $\ln N + 1.6\ln D$ = constant; and Kira et al. (1953; also Yoda et al.., 1963) advocated the self-thinning line which relates limiting stocking (*N*) to plant mass (*V*): $\ln N + \tfrac{2}{3}\ln V$ = constant. Attempts to define these limiting conditions are often demanding of data, may require long-term monitoring of undisturbed stands, and involve subjective decisions to select data close to the assumed limit.

Goulding (1972) advocated a hyperbolic relationship between diameter, height and stocking: $D = (\beta_1 H^{\beta_2} N + \beta_3 H^{\beta_4})^{-1}$. Sterba (1975, 1987) extended this work and showed how to calculate the theoretical maximum basal area. Sterba's work has formed the basis for some tree growth models (e.g., Hasenauer, 2006; Skovsgaard, 1997), but calibrating these models can be demanding of data and may not be possible in situations where data are scarce. Such situations commonly occur with tree plantings for environmental remediation, for non-timber forest products and in other situations where afforestation involves new species or non-traditional sites. This paper seeks to offer a robust way to provide an approximate but reliable growth projection in such situations where empirical data are scarce and resources preclude calibration of physiological growth models.

**Materials and Methods**

The relationship, $D = \beta (H-1.3)/\text{Ln}(N)$ is a straight line passing through the origin and relates the mean diameter of forest trees ($D$, diameter at breast height, 1.3 m above ground) to the stand height ($H$, average height of the largest trees in a stand) divided by the logarithm of stand density ($N$, number of trees per hectare). This one-parameter model is inherently robust and simple to estimate, and may be useful in data-scarce situations where established approaches cannot be calibrated. Many researchers have examined the diameter-height-density relationship for individual trees within stands (e.g., Curtis, 1967; Lynch et al., 2007; Mohler et al., 1978; Niklas et al., 2003; Rio et al., 2001; Temesgen and Gadow, 2004; Vanclay and Henry, 1988; Zeide, 1995; Zeide and Vanderschaaf, 2002), but this particular relationship has apparently not been reported previously. This relationship has been observed empirically for several sites and species, and is consistent with expected natural stand dynamics. Stand height $H$ reflects past performance and is extensively used as a measure of site (Skovsgaard and Vanclay, 2008), is correlated with both crown width and root length (Hemery et al., 2005; Kajimoto et al., 2007; Niklas and Spatz, 2006; Peichl and Arain, 2007; Usoltsev and Vanclay, 1995) and thus reflects the resources potentially available to a tree. Since $N$ reflects competition for these resources, the interaction $H/\text{Ln}(N)$ may provide a useful expression of growth potential of individual trees. The offset $(H-1.3)/\text{Ln}(N)$ ensures that the trend passes through the origin for trees approaching breast height and thus with zero diameter. $D$ reflects the extent to which individual trees have been able to capitalise on this potential. The slope $\beta$ should reflect the efficiency of resource use, and may be species- and site-specific.

There are several ways in which the relationship between $D$, $H$ and $N$ may be explored. Some more obvious options include $H/D=a+b\text{Ln}(N)$, $D = bH^c N^e$, and $D = (\beta_1 H^{\beta_2} N + \beta_3 H^{\beta_4})^{-1}$, but the proposed formulation $D = \beta (H-1.3)/\text{Ln}(N)$ offers the robustness of a single parameter to estimate, and provides for a distribution of residuals consistent with conventional statistical assumptions.

The relationship $D = \beta (H-1.3)/\text{Ln}(N)$ has been tested using data collated through an Australia-wide project under the auspices of the Standing Committee on Forestry of the Ministerial Council on Forestry, Fisheries and Aquaculture, through its Research Working Group 2 (Mattay and West, 1994), as well as with other published data. Table 1 offers a summary of data sets examined. These data include permanent sample plots monitored for up to 60 years, span stem sizes up to 80 cm in diameter and 65 m in height, represent stand density ranging from 30 to 84000 stems/ha, and sample a diverse range of locations (Table 1), thus offering a rigorous test of the relationship.

Estimates of height and diameter used in these analyses generally refer to predominant height (expected height of tallest 100 stems/ha) and arithmetic mean diameter, but in some instances refer to quadratic mean diameter $((\Sigma D^2/N)^{0.5})$, top height (expected height of thickest 100 stems/ha) or other local variations (Sharma et al., 2002). These alternative measures change the slope $\beta$, but do not alter the underlying relationship between diameter, height and stocking. The relationship appears to apply equally well to arithmetic and quadratic mean diameters, and to top, predominant and average height, with only minor changes in the slope $\beta$.

**Results**

Figure 1 illustrates the relationship $D = \beta (H-1.3)/\text{Ln}(N)$ for 626 measurements in 97 plots of even-aged *Eucalyptus pilularis* in Queensland and New South Wales, Australia. The dotted line illustrates the average trend with $\beta=5.6$ (s.e.=0.026), while the individual plot estimates vary between 4.1 and 7.8. A straight line through the origin offers an adequate ($P<0.001$) and sufficient (test for lack of fit: $P>0.06$) fit to these data. The illustrated data span a range of density 80-7180/ha, of mean diameter 2-67 cm, and of stand height 4-54 m, indicating the stability of the relationship. These plots were established from 1931, span a time interval of up to 55 years, and were measured up to 31 times during that period. Five individual plot trajectories have been highlighted to illustrate the temporal development of these forest stands.

This relationship does not imply a constant relationship between the height and diameter of individual trees, because the observed trend applies to the mean diameter and the stand height, not to the dimensions of individual trees. Figure 2 illustrates the heights and diameters of 44 trees in a 0.16 ha plot of *Eucalyptus pilularis* in Queensland, and demonstrates how individual heights and diameters exhibit allometric relationship while contributing toward a linear trend $D = \beta (H-1.3)/\text{Ln}(N)$ at the stand level.

Similar trends emerge for other regrowth stands of eucalypts (e.g., *E. grandis* in Fig. 3; $\beta=5.1$, s.e.=0.028, adequate $P<0.001$, sufficient $P>0.3$). This relationship applies moderately well to pooled data from many sites, but performs better for individual stands – once a stand has established a trajectory, it tends to maintain the same trajectory long-term, relatively unaffected by thinning and other silvicultural intervention.

The relationship is not confined to the genus *Eucalyptus*, but appears to hold for a wide range of sites and species (Table 1). Figure 4 shows this relationship for *Tectona grandis* forests in Myanmar with data from Laurie and Ram (1939), illustrating how a wide range of stands tend to follow similar trajectories ($\beta=6.2$, s.e.=0.024, adequate $P<0.001$, sufficient $P>0.2$).

Figures 1-4 relate to indigenous species in both plantations and naturally-regenerated stands, but the relationship also applies to plantations of exotic species. Figure 5 illustrates the development of a *Pinus radiata* plantation in South Australia, based on 25 annual measurements of a 10-35 year-old plantation ($\beta=7.8$, s.e.=0.045, $P<0.001$). Solid points indicate the post-thinning condition after five thinnings, each of which reduced stocking by approximately 40%, gradually reducing the density from 1800/ha to 167/ha, while maintaining the stand basal area within the range 21-37 m$^2$/ha. Figure 5 illustrates the tendency to maintain a relatively constant relationship, with any perturbations due to conventional thinning small and transient. With these data, the simple linear relationship is adequate ($P<0.001$), but there is a small but significant trend with stocking ($P=0.03$) indicating that the simple relationship tends to overestimate in stands with stocking below 180 stems/ha (indicated in Fig. 5 with grey-filled points).

More extreme thinning regimes may cause larger perturbations, but the overall trend tends to remain, even for extreme treatments. Figure 6 illustrates the 60-year trajectory of four *Picea abies* plots in Denmark, some of which were thinned frequently, while one plot remained unthinned and at high density throughout the period of monitoring. The thinning treatments involved reductions from 5428-2017 stems/ha in 3 thinnings (A), from 4435-275/ha in 16 thinnings (B), from 1113-160/ha in 22 thinnings (D), and from 733-126/ha in 14 thinnings (L). Despite these contrasting treatments, the four plots tend to maintain a stable trajectory (Fig. 6) that is adequate ($P<0.001$) and sufficient ($P=0.06$), except for a tendency to underestimate when stocking falls below 150 stems/ha (indicated in Fig. 6 with grey-filled points).

Density variations due to initial spacing of plantings have also little effect on the relationship. Figure 7 illustrates the growth trajectories of *Araucaria cunninghamii* trees planted in a mixed-species Nelder (1962) wheel in Queensland, Australia. This planting included eight densities ranging from 42-3580 stems/ha (Lamb and Borschmann, 1998). Despite the wide range of stocking, the tendency to follow a stable trajectory remains evident ($\beta=10.5$, s.e.=0.13, $P<0.001$), notwithstanding some tendency towards a quadratic relationship ($P>0.2$). However, in Figure 7, it is the trees at high stocking (>2000/ha), rather than those with low stocking, that depart from the trend. This may be partly due to tree geometry (lean and branching), which, in a Nelder wheel, can change the effective spacing. Figure 8 includes data from both species (*A. cunninghamii* and *Flindersia brayleyana*) in the Nelder wheel, illustrating the tendency for intimate mixtures of tree species to exhibit similar slopes. This aspect warrants further investigation.

**Discussion**

This relationship has apparently not been recognised previously, but appears to hold for a wide range of species and conditions (Table 1). The one-parameter nature of this relationship makes it easy to calibrate, even when data are limiting. Empirical studies suggest that the slope $\beta$ varies between 5 and 10 for different sites and species (Table 1). Species planted as polycultures appear to exhibit similar values for $\beta$ (Fig. 8, Table 1), suggesting that $\beta$ may be site-dependent rather than species-dependent, but further research is warranted to confirm this observation.

The relationship has several potential applications in forest resource assessment and forest management. $N$ and $H$ are easily determined via remote sensing (e.g., lidar), and the relationship allows diameters, and thus volume and sequestered carbon to be inferred remotely. Because the relationship is stable over time, it may be used to make robust forecasts of future growth. Growth modellers may use the relationship to evaluate existing models (Vanclay and Skovsgaard, 1997), and as an additional constraint when simultaneously estimating relationships describing tree growth.

The relationship is not universal. It appears to hold for stands with 200-2000 stems/ha, but may fail for extremes of stocking. It appears to hold for a wide range of thinning practices, provided that sufficient time elapses between thinnings for a new equilibrium to be established. These caveats mean that the relationship $D = \beta (H-1.3)/\mathrm{Ln}(N)$ should be viewed as a first approximation rather than as a definitive model. There may be circumstances where there is a small but significant intercept, where a quadratic term is significant, or where an alternative formulation (e.g., $D = bH^c N^e$) may provide a significantly better fit (Zeide, pers. comm., 2008), but these alternatives lack the robust utility of the simple one-parameter model, especially in data-poor situations.

**Conclusion**

In even-aged stands, the mean diameter of forest trees ($D$, diameter at breast height, 1.3 m above ground) tends to remain proportional to the stand height ($H$, average height of the largest trees in a stand) divided by the logarithm of stand density ($N$, number of trees per hectare): $d = \beta (H-1.3)/\mathrm{Ln}(N)$. Thinning causes a relatively small and temporary perturbation to $\beta$, the magnitude and duration of which depends on the nature of the thinning.

**Acknowledgements**

My colleagues David Lamb (University of Queensland, Australia), Jerry Leech (Woods and Forests Department, South Australia), JP Skovsgaard (Danish Centre for Forest, Landscape and Planning) and Vladimir Usoltsev (Forest Institute, Russian Academy of Sciences) kindly provided data for use in this analysis. The bulk of the data was kindly provided by Julian Mattay, custodian for Research Working Group 2 of the Standing Committee on Forestry of the Ministerial Council on Forestry, Fisheries and Aquaculture. Thanks are also due to Boris Zeide for helpful comments on the draft manuscript.

**Table 1**. Estimates of β for even-aged plantings.

| Species | Location | β | $R^2$ | N | Max D (cm) | Max H (m) | Min N (ha$^{-1}$) | Max N (ha$^{-1}$) | Reference |
|---|---|---|---|---|---|---|---|---|---|
| *Eucalyptus regnans* | Tasmania‡ | 4.51 | 0.88 | 279 | 76.6 | 62.8 | 61 | 5460 | Mattay & West 1994 |
| *E. pellita*[1] | Queensland‡ | 5.32 | 0.89 | 32 | 30.3 | 28.8 | 107 | 476 | Bristow et al. 2006 |
| *Acacia peregrina*[1] | Queensland‡ | 5.47 | 0.77 | 32 | 20.7 | 20.3 | 71 | 571 | Bristow et al. 2006 |
| *E. grandis* | Queensland‡ | 5.58 | 0.95 | 202 | 66.9 | 55.6 | 100 | 1820 | Mattay & West 1994 |
| *E. diversicolor* | Western Australia | 5.72 | 0.74 | 104 | 76.4 | 65.0 | 320 | 3720 | Mattay & West 1994 |
| *Pinus radiata* | ACT, Australia | 5.99 | 0.97 | 359 | 30.5 | 36.7 | 537 | 2990 | Snowdon & Benson 1992 |
| *Tectona grandis* | Myanmar | 6.09 | 0.94 | 642 | 65.0 | 35.7 | 69 | 2098 | Laurie & Ram 1939 |
| *E. pilularis* | Queensland‡ | 6.23 | 0.91 | 626 | 66.6 | 54.1 | 80 | 7180 | Mattay & West 1994 |
| *Betula pendula* | Kazakhstan | 6.54 | 0.94 | 60 | 22.0 | 23.4 | 340 | 33100 | Usoltsev (pers comm 1999) |
| *E. obliqua* | Tasmania‡ | 6.56 | 0.79 | 473 | 76.2 | 55.5 | 40 | 7663 | Mattay & West 1994 |
| *E. delegatensis* | Tasmania‡ | 6.70 | 0.95 | 115 | 53.7 | 50.4 | 140 | 3860 | Mattay & West 1994 |
| *Pinus radiata* | South Australia | 6.99 | 0.93 | 3535 | 76.4 | 44.6 | 72 | 2899 | Leech (pers comm 1994) |
| *Pinus sylvestris* | Kazakhstan | 7.51 | 0.95 | 95 | 28.8 | 25.1 | 400 | 84000 | Usoltsev (pers comm 1999) |
| *Picea sitchensis* | Denmark | 7.84† | 0.94 | 998 | 44.2 | 26.8 | 30 | 7759 | Vanclay et al. 1995 |
| *Araucaria cunninghamii*[2] | Queensland‡ | 8.83 | 0.96 | 104 | 29.6 | 17.1 | 42 | 3580 | Lamb & Borschmann 1998, Vanclay 2006 |
| *Flindersia brayleana*[2] | Queensland‡ | 9.32 | 0.94 | 108 | 30.2 | 17.3 | 42 | 3580 | Lamb & Borschmann 1998, Vanclay 2006 |

† This study used quadratic mean diameter $(\Sigma D^2/N)^{0.5}$, which results in a higher slope than arithmetic mean D.
‡ Australia.
1,2 These species planted as a mixture (1: Bristow et al. 2006; 2: Lamb and Borschmann 1998).

**Figures**

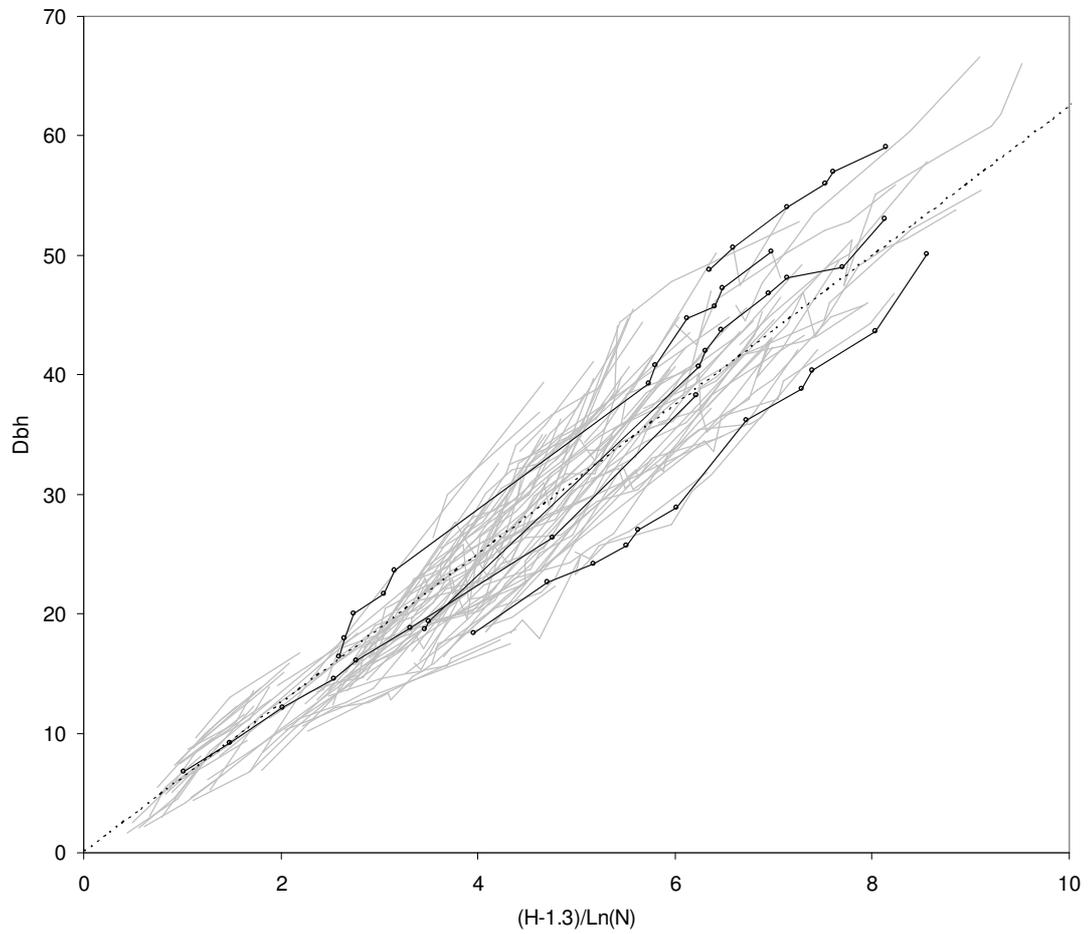

**Figure 1**. The relationship $D = \beta\,(H\text{-}1.3)/\text{Ln}(N)$ illustrated for 626 measurements in 97 plots of even-aged *Eucalyptus pilularis* in Queensland and New South Wales, Australia, with 5 trajectories illustrating the temporal development of forest stands (Data from Mattay and West 1994).

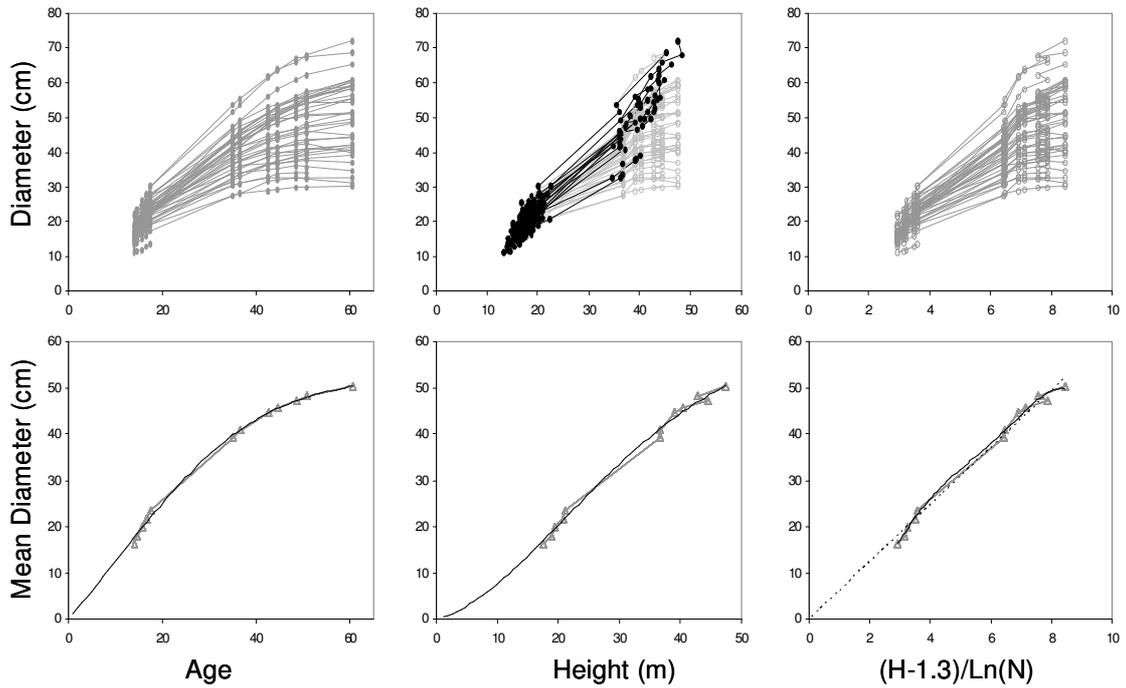

**Figure 2.** Individual (top) and stand average (bottom) relationships between diameter and height. Graphs illustrate diameter versus age (left column), versus height (centre column), and versus (H-1.3)/Ln(N). The bottom row includes a quartic polynominal (black line) and the linear trend (dotted line, bottom right). The diameter-height graph (top centre) includes individual tree heights (black) and predominant height (grey, tallest 50 trees/ha) for trees not measured for height.

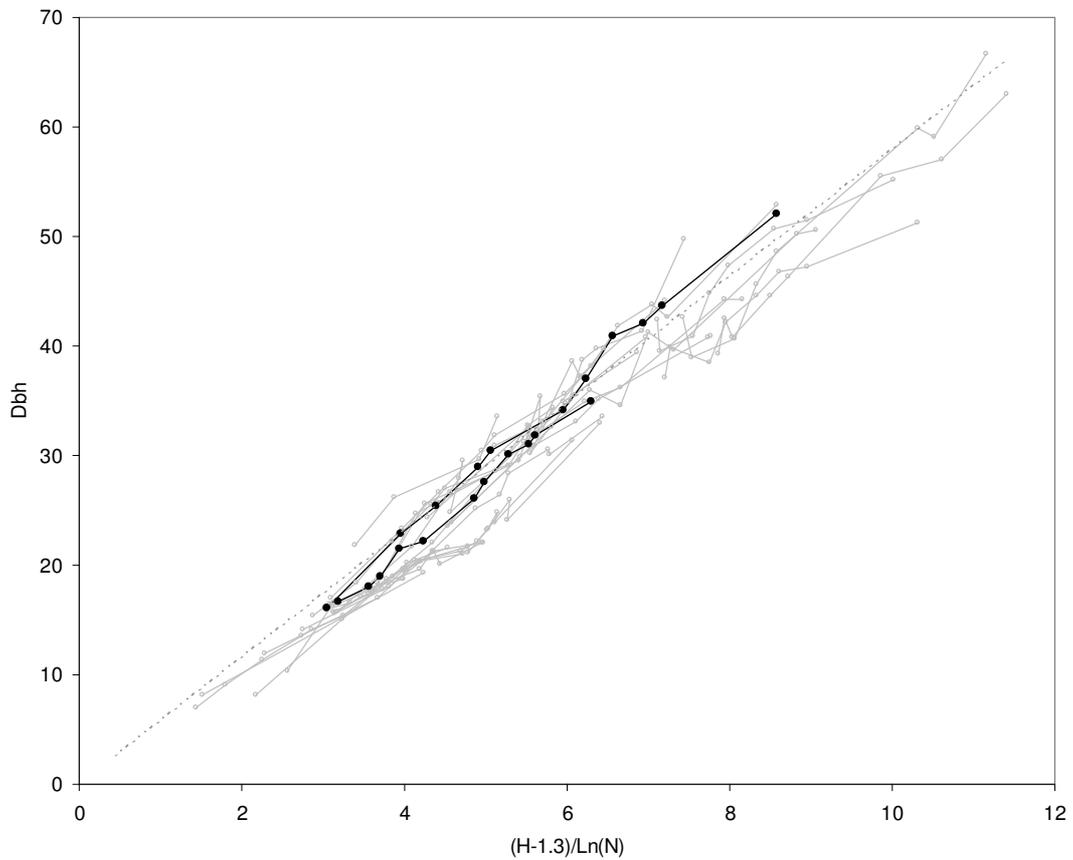

**Figure 3**. The relationship $D = \beta (H\text{-}1.3)/\text{Ln}(N)$ illustrated for 202 measurements in 25 plots of even-aged *Eucalyptus grandis* in Queensland, Australia, with 2 trajectories illustrating the temporal development of individual forest stands. Data span the years 1941-87, and a range of stocking 80-1075/ha, of mean diameter 1-72 cm, and of stand height 1-56 m (Data from Mattay and West 1994).

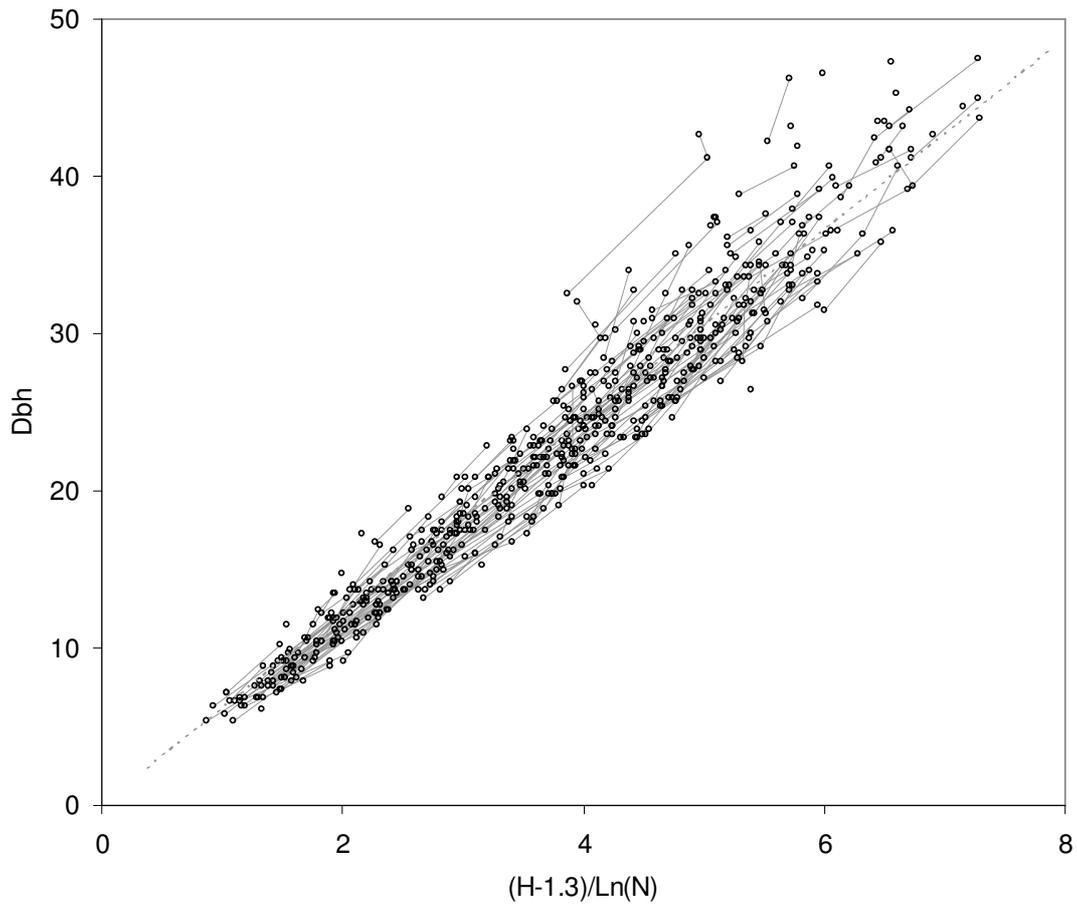

**Figure 4**. Development of *Tectona grandis* stands in Myanmar. Data (Laurie and Ram 1939) represent 642 measurements of 327 plots, spanning stands ranging from 3-72 years in age, 7-35 m in height, and 4-48 cm in diameter.

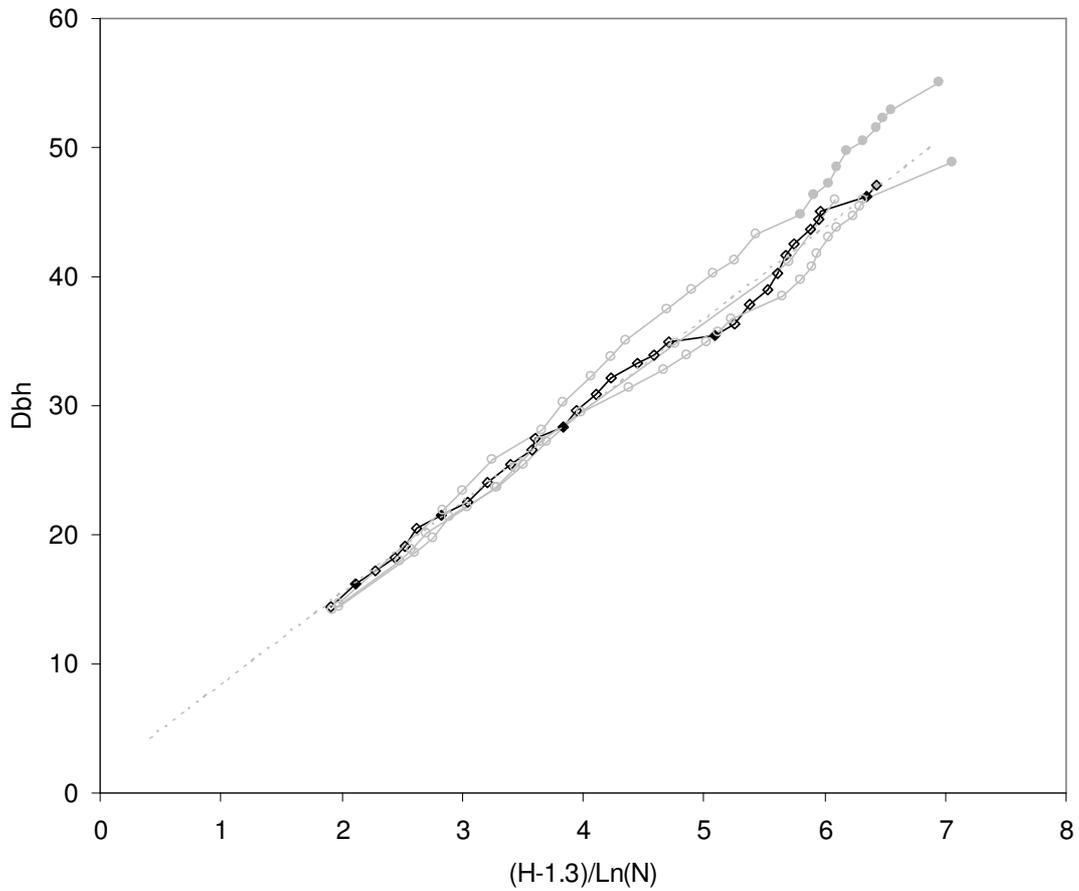

**Figure 5**. Development of *Pinus radiata* in South Australia from age 10 to 35 years. Black line illustrates the development of stand thinned on 5 occasions (solid black symbols ♦) from 1800 to 167 stems/ha. Grey lines illustrate three similar stands thinned from 1800 to 230 (middle line) and 140 stems/ha (outermost lines), and points filled grey indicate stocking <180 stems/ha. Data courtesy of Dr JW Leech, Woods and Forests Department, South Australia.

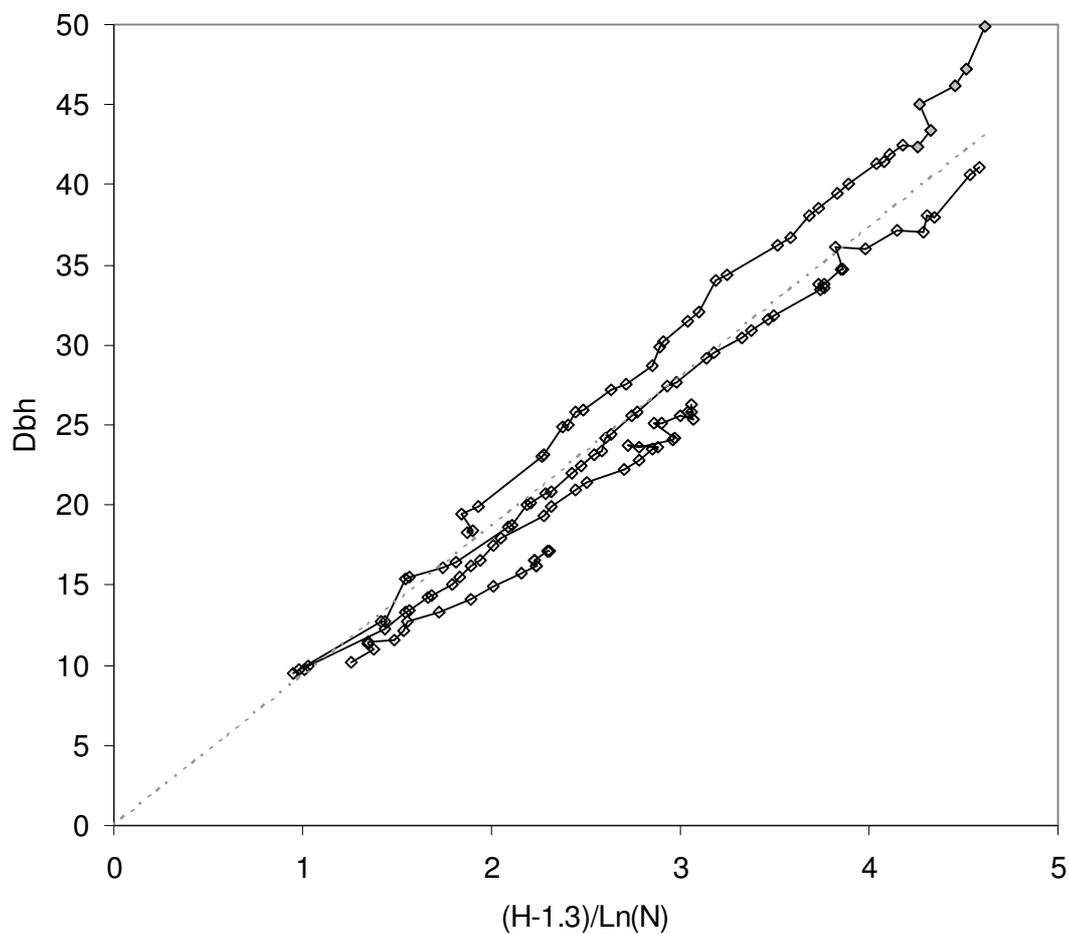

**Figure 6**. The development of *Picea abies* in Denmark illustrated with 60-year trajectories from stands thinned to four prescriptions (*A, B, D, L*). The *A* treatment (lowest line) had 3 thinnings progressively reducing density from 5428 to 2017 stems/ha. The L treatment (uppermost line) reduced density from 733 to 126 stems/ha gradually in 14 thinnings. Grey-filled points indicate stocking <150 stems/ha. Data courtesy of Prof J.P. Skovsgaard, Danish Centre for Forest, Landscape and Planning.

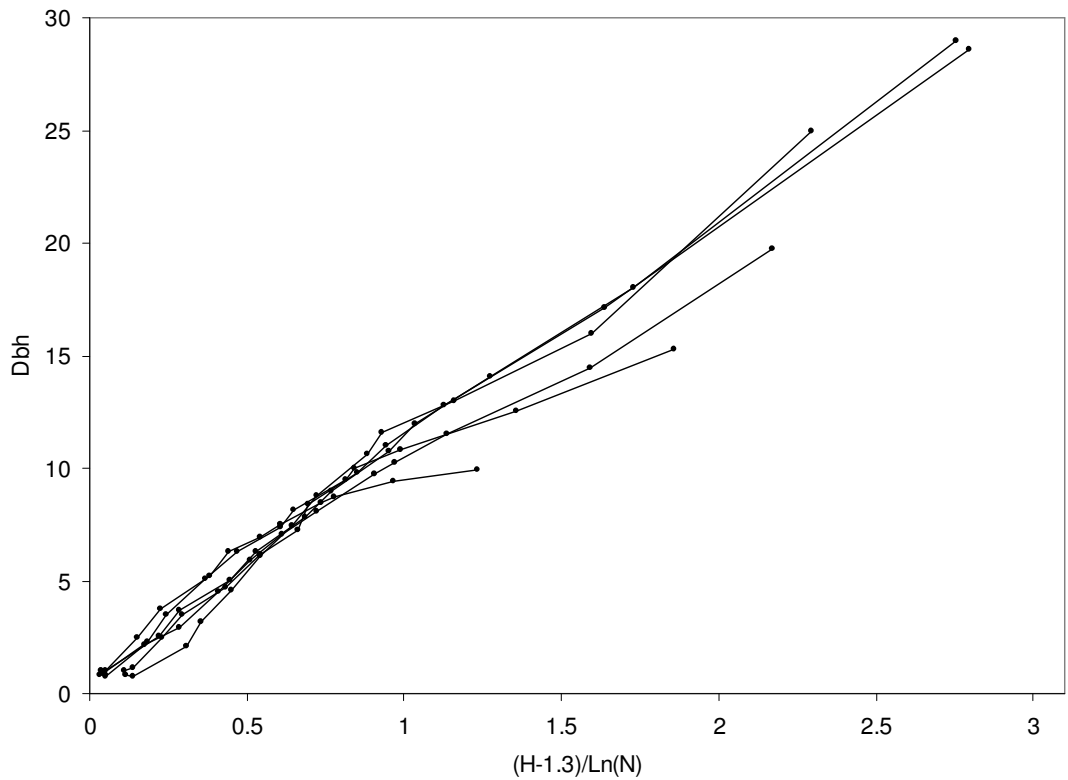

**Figure 7**. Temporal development of *Araucaria cunninghamii* in a Nelder wheel in southeast Queensland (Lamb and Borschmann, 1998), representing stand densities ranging from 42-3580 stems/ha. Stems vary from 1-30 cm diameter and 1-17 m height.

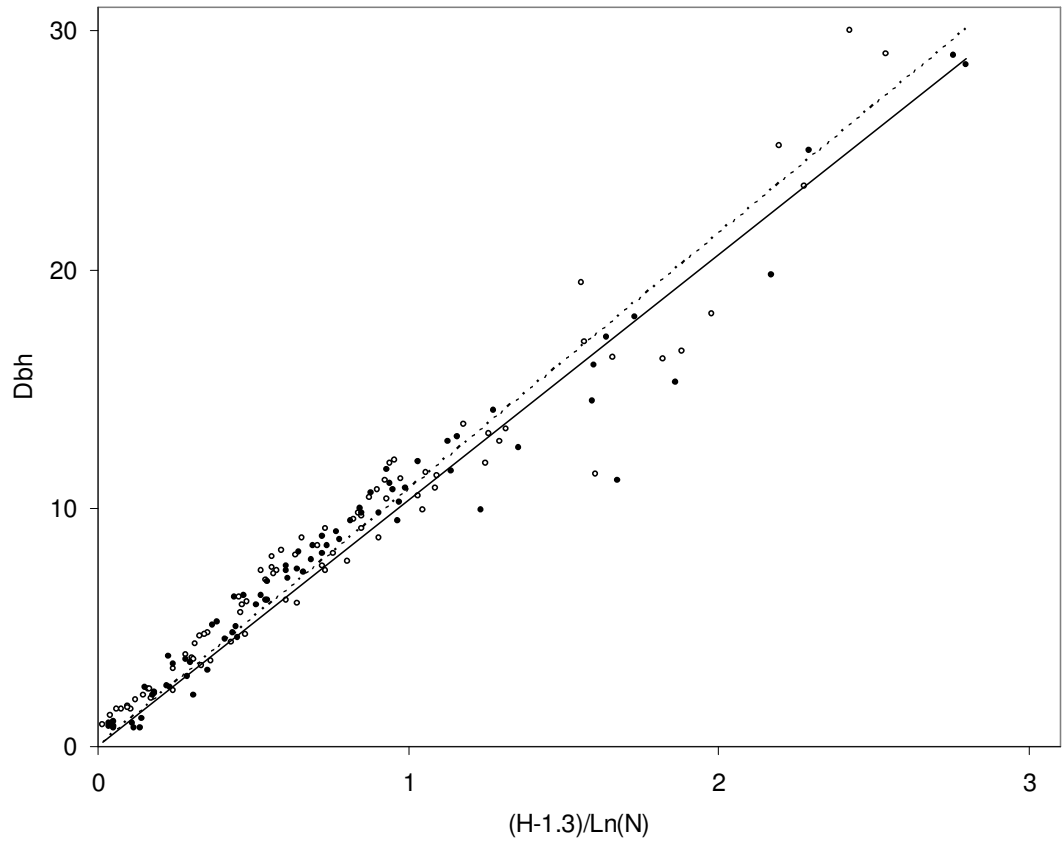

**Figure 8**. Data from a mixed planting of the coniferous *Araucaria cunninghamii* (●, solid line) and the broad-leaved *Flindersia brayleyana* (○, dotted line) in south-east Queensland (Lamb and Borschmann, 1998), illustrating that species with different growth habits can exhibit similar slopes.